\documentclass[prd,showpacs,showkeys,preprint,nofootinbib]{revtex4}
\usepackage{color}
\usepackage{amsmath}
\usepackage{amssymb}
\usepackage{yfonts}[1988/10/03]
\usepackage{graphicx}
\usepackage{subfig}
\newcommand {\beq}{\begin{equation}}
\newcommand {\eeq}{\end{equation}}
\newcommand {\beqa}{\begin{eqnarray}}
\newcommand {\eeqa}{\end{eqnarray}}

\begin{document}
	{\textsf{\today}
\title{ Gravitational Waves in a Closed Spacetime via Deviation Equation }
\author{Jafar Khodagholizadeh $^{1}$\footnote{gholizadeh@ipm.ir}, Amir H. Abbassi $^{2}$\footnote{ahabbasi@modares.ac.ir}, Ali Vahedi$^{3}$\footnote{vahedi@ipm.ir } and  K. Babaei  $^{1}$\footnote{komeil.babaee@gmail.com }}
\affiliation{$^{1}$ Farhangian University, P.O. Box 11876-13311,  Tehran,  Iran.\\
$^{2}$
Department of Physics, School of Sciences, Tarbiat Modares University, P.O. Box 14155-4838, Tehran, Iran. \\
$^{3}$
Faculty of Physics, Kharazmi University, P.O. Box 15614,  Tehran, Iran.}

\begin{abstract}
Within the closed universe, we obtain the amplitude and frequency of gravitational waves in the terms of discrete wave numbers, wave propagation time, and cosmological constant using the deviation equation in the first-order perturbed metric. We demonstrate that the cosmological constant effect on GWs is only seen in the early universe. Also, by considering the time evolution of a gravitational wave in a closed spacetime, we investigate its effect on a circle of nearby massless particles, which will be compared with this case in the flat spacetime. Expanding the universe has effective damping on GWs; thus, we suggest it can be used as a tool to characterize the large-scale curvature of the universe.   
\end{abstract}
%\preprint{***}
\maketitle
\section{Introduction} 
 The recent detection of gravitational waves (GWs) by LIGO-Virgo has opened a new window into the universe \cite{Ligo}. It was a first hint of the great potential of GWs detection for the universe exploration and understanding. For the standard theoretical analysis, GWs rely on the split of the spacetime metrics $g_{\mu \nu} $ into small perturbation in addition to the background metric. In general relativity, the curvature of the background is the fundamental field, characterizing gravity, which is not required to be the Minkowski spacetime. While our universe appears to be spatially flat approximately at the accuracy of $0.5\%$, the data do not entirely rule out the case of $K=1 $ \cite{Planck, Planck3, Li}.\\ Recently, the Planck legacy 2018 CMB power spectra provide statistically significant indications for a closed universe \cite{Pl18}. The curvature tension is considered evidence for closed spacetime in\cite{Will}. Also, a positive curvature is marginally suggested by the ages of the oldest stars\cite{Schla, Bond}. In addition, Planck 2018 CMB lensing \cite{lensin} and baryon acoustic oscillations \cite{14, 15, 16} suggest a flat universe are quantitatively inconsistent at $ 2.5 $ to $ 3\sigma $  with CMB data.  Hence, by choosing the maximally extended de-Sitter metric (K = 1) as the unperturbed background, we studied the propagation of gravitational waves \cite{Gholizadeh} and its damping by neutrinos in closed spacetime \cite{Gholizadeh1, Gholizadeh2}.\\ As before, many authors have considered the de Sitter background to study GWs; M.Shibata et al. investigated the dynamical evolution of axisymmetric gravitational waves in asymptotically de-Sitter spacetime \cite{Shibata} and found that, if the mass of gravitational waves is larger than the critical value, $ M_{crit,}=(3\sqrt{\Lambda})^{-1} $ the formation of black holes will be prevented, even in the presence of highly non-linear and localized gravitational waves. Within Bondi-Saches formalism, some properties of GWs about de Sitter background have been investigated\cite{Bishop}. Also, there has been an analysis on the behavior of the gravitational (electromagnetic) waves with the simplicity of the conformal transformation around de-Sitter background \cite{Bicak1, Bicak2}. \\
The GW's propagation has been investigated in an asymptotically de-Sitter spacetime by expanding the perturbation around the Minkowski spacetime in the presence of the cosmological constant, which at once $ \Lambda $ is an additional source ( de-Donder gauge) and after is a gauge ($ \Lambda $-gauge) and then concluded that the cosmological constant,  $ \Lambda $ impedes the detection of gravitational waves at the distance larger than $ L_{crit}\approx r_{\Lambda}^{2} $  with $r_{\Lambda}=\frac{1}{\sqrt{\Lambda}}$ \cite{Ivan1}.\\ 
The effect of the cosmological constant on the gravitational waves as a background perturbation, corresponding to the curvature; $ \eta_{\mu\nu} + h_{\mu\nu}^{\Lambda}$,  has been investigated; the wave-like perturbations can absolutely have a detectable impact on pulsar timing arrays and these waves are modified both in the phase and the amplitude \cite{Jose}. The linearized gravitational waves in de Sitter spacetime are analyzed in the presence of a positive cosmological constant to obtain the quadrupole formula and the theory of gravitational radiation\cite{Abhay, Abhay1, Abhay2}. Another use of the cosmological constant, $ \Lambda $ is in the energy-momentum pseudo-tensor of the gravity and the result can be formulated in the form of the critical distance proportional to $r_{\Lambda}=\frac{1}{\sqrt{\Lambda}}$, the frequency and the amplitude of GWs, as well as the distance of the source from the detector \cite{Nowa}.\\
As mentioned in \cite{Gholizadeh}, we found the propagation of the gravitational waves in the closed background which depends on the wave numbers only. Thus far, the frequency and amplitude of gravitational waves have not been achieved; its effect on the particle's ring in the de Sitter background has not been investigated.\\
The current paper is organized as follows: In Section II, we review the geodesic deviation equation (GDE) in a closed spacetime and its lowest order solution will be found in the next section with its effect on the nearby massless particles. Also, we provide the amplitude and frequency of GWs. Then, in Section IV, using the general solution of gravitational waves equation in closed spacetime, we investigate the passing of these waves on the particles ring and study the presence of an effectively closed background. Section V contains a discussion on more significant effects, quantitative differences, or sensitivity of the detector. Finally, we compare the results with those of the flat case.
\section{ Geodesic Deviation Equation on Closed Spacetime }
To study the physical effects due to gravitational waves, the motion of test particles in the presence of this wave is considered. Although these test particles are massless, in Newtonian McVittie's background, the propagation of gravitation waves has been investigated in the presence of point mass \cite{Antoniou}. Also, to obtain the coordinate independent measurement, the relative motion of the nearby particles is described by the geodesic deviation equation (GDE). Therefore, there are some nearby particles with four-velocity described by a single vector field $ U^{\mu}(x)$ and a separation vector $ S^{\mu} $. The geodesic deviation equation is \cite{6}:
\begin{equation}
\dfrac{D^{2}}{D\tau^{2}}S^{\mu}=R^{\mu}_{~\rho\sigma \nu}U^{\rho}U^{\sigma}S^{\nu}
\end{equation}
as we work in the first order metric perturbation in $h_{\mu\nu}$, e.g. $ g_{\mu\nu}=\bar g_{\mu\nu}+ h_{\mu\nu}$ where $ \bar g_{\mu\nu}$ is the metric of background}. If we take test particles to be moving slowly, the four-velocity can be expressed as a unit vector in the time direction, plus corrections of order $h_{\mu\nu}$ and higher. The Reimann tensor already is in the first order, so the correction to $ U^{\nu} $ may be ignored. Thus, we write:
\begin{equation}
U^{\rho}=(1,0,0,0)
\end{equation}
Therefore, we need to calculate $ R^{\mu}_{00\nu} $. In general, the background is flat (or Minkowski) spacetime that can be found in any textbook of General Relativity or Cosmology (e.g., in \cite{6}). This paper considers the maximally extended de Sitter spacetime as a background metric, the components of which in the Cartesian coordinates are \cite{Weinberg}
\begin{equation}
\bar{g}_{00}=-1~~~~~ ,~~~~ \bar{g}_{0i}=0 ~~~~ and~~~~ \bar{g}_{ij}=a^{2}(t)\tilde{g}_{ij}= a^{2}(t)(\delta_{ij}+K\dfrac{x^{i}x^{j}}{1-K x^{2}})
\end{equation}
where $i$ and $j$ run over the values 1, 2 and 3, and $ x^{0}\equiv t $ is the time coordinate  in our units, with the speed of light equal to unity. Also, $x^{i}$ are the quasi-Cartesian coordinates,  $K$ is curvature constant, and $ a(t)=\alpha \cosh(\dfrac{t}{\alpha}) $ is the scale factor in de Sitter spacetime with $ \alpha=\sqrt{\dfrac{3}{\Lambda}} $ \footnote{ Note: We assume that the radius of de Sitter spacetime ($l$) is unity. So, what we consider as cosmological constant is $\Lambda l^2$, which is a dimensionless value. If we consider $l\approx 10^{26} m$ as the radius of our observable universe, our dimensionless $\Lambda$ is in the unit of $10^{-52} m^{-2}$.}. The Riemann tensor with the first order perturbation is
\begin{equation}
R^{\rho}_{\mu\nu\lambda}=\bar{R}^{\rho}_{\mu\nu\lambda}+\delta R^{\rho}_{\mu \nu\lambda}
\end{equation}
where $ \bar{R}^{\rho}_{\mu\nu\lambda}$ is the Riemann tensor of the background. The $R^{i}_{~00j} $ is: 
\begin{equation}
R^{i}_{~00j}=\dfrac{\ddot{a}}{a}\delta^{i}_{\nu}+\dfrac{1}{2}\tilde{g}^{i k}\ddot{D}_{k j}+\dfrac{\dot{a}}{a}\tilde{g}^{ik}\dot{D}_{kj}
\end{equation}
where $D_{ij}= a^{-2}(t) h_{ij}$ and it is the solution for gravitational waves equation in curved spacetime \cite{Gholizadeh}:
\begin{equation}\label{41}
\nabla^{2} D_{ij}-a^{2}\ddot{D}_{ij}-3 a \dot{a}\dot{D}_{ij} -2KD_{ij}=0
\end{equation}
The last term is the presence of non-zero curved background. To be more specific and without losing the generality, we chose the solution of equation (\ref{41}) as a move in the z-direction in closed spacetime ($K=1$), e.g. for $ + $ mode from the traceless and transverse conditions in the closed background, we have\cite{Gholizadeh}
\begin{equation} \label{29}
D^{i}_{+j}=\frac{D_{+}(z,\tau)}{\sqrt{1-X^2}}\left(\begin{array}{ccc}
\dfrac{1}{(1-x^{2}-z^{2})}& 0  &\frac{xz}{(1-z^{2})(1-x^{2}-z^2)}\cr 0& -\dfrac{1}{1-y^{2}-z^{2}}& \dfrac{-yz}{(1-z^{2})(1-y^{2}-z^{2})}\cr \dfrac{xz}{(1-z^{2})(1-x^{2}-z^{2})}&\dfrac{-yz}{(1-z^{2})(1-y^{2}-z^{2})}&\dfrac{z^{2}(x^{2}-y^{2})}{(1-z^{2})(1-y^{2}-z^{2})(1-y^{2}-z^{2})}\end{array}\right).
\end{equation}
where $ X^{2}=x^{2}+y^{2}+z^{2} $ and $ D_{+}(z,t) $ is:
\begin{eqnarray}
{D}_{+}(z,\tau)=\frac{(1-z^2)}{1-n^2}(\cos\tau \mp in\sin\tau)\left\{
\begin{array}{c}
exp[\pm in(\arccos z+\tau)]\cr
exp[\pm in (\arccos z-\tau)].
\end{array}\right.
\end{eqnarray}
where $ n $ is the wave number and is an integer which should be discrete (our attempt to find periodic waves requires the wave number to be integer. We will have the same result for $\times$ mode; but, only its matrix will be different and ${D}_{+}(z,\tau)={D}_{\times}(z,\tau)$ (For more details, please see \cite{Gholizadeh}).  

It can be seen that the time propagation of GWs in the closed spacetime is not simply a plane wave.  Also, it has all properties of the transverse-traceless condition (please see Appendix A). Therefore, the deviation equation in the closed background becomes \cite{Gholizadeh}
\begin{equation} \label{deviation}
\dfrac{d^{2}}{d\tau^{2}}S^{i}=\frac{\Lambda}{3}S^{i}+\dfrac{\dot{a}}{a}S^{j}\frac{d}{d\tau} D^{i}_{~j}+\dfrac{1}{2}S^{j}\dfrac{d^{2}}{d\tau^{2}}D^{i}_{~j}
\end{equation}
As is clear from the above equation, the relative acceleration between the two neighboring geodesics is proportional to the tensor perturbation, the cosmological constant, and the expansion parameter.
The first term in Eq.\eqref{deviation} provides exponential expansion, which is a general characteristic of de Sitter spacetime. The $H(t)=\dfrac{\dot{a}(t)}{a(t)}$ is the Hubble parameter, expressing the effect of expanding the universe.
By considering all terms on the right-hand side of Eq.\eqref{deviation}, the acceleration of neighboring geodesics will be interpreted physically as a gravitational tidal force. However, in the next section, we will find the general solution of the deviation equation in closed spacetime and, then, investigate only the effect of closed gravitational waves on closely spaced particles.
\section{ General Solution of Deviation Equation and its Effect}
To find complete solution of deviation equation for slowly moving particle, or the lowest order solution, we could write Eq. (\ref{deviation}) as:
\begin{equation}
\Big(\dfrac{d^{2}}{d\tau^{2}}-\frac{\Lambda}{3}\Big)S^{i}=S^{j} L_\tau D^{i}_{~j}
\end{equation}
where $L_\tau=\dfrac{\dot{a}}{a}\dfrac{d}{d\tau} +\dfrac{1}{2}\dfrac{d^{2}}{d\tau^{2}}$ is a linear differential operator. Also, as mentioned, $ D_{+}(z,\tau)= D_{\times}(z,\tau)$,  we consider the mixed tensor as $D^{i}_{~j}=\delta^{i}_{~j} D_{+, \times}(z,\tau)$. Then,
\begin{equation}\label{eq:2}
\dfrac{d^{2}}{d\tau^{2}}S^{i}=\frac{\Lambda }{3} f(\tau,\vec{x},\Lambda) S^{i} 
\end{equation}
where $f(\tau,\vec{x},\Lambda)$ is defined as follows:
\begin{align}
f(\tau, \vec{x},\Lambda)=1+\dfrac{3}{\Lambda} L_\tau D_{+, \times}(z,\tau)
\end{align}
For the wave traveling in the z-direction, only $ S^{1} $ and $ S^{2} $ will be affected. For slow-moving particles, we have $ \tau=x^{0}=t $. Therefore, to the lowest order for $ " + " $ modes, the solutions are:  
\begin{eqnarray}\label{12}
S^{1}(t, n, \Lambda)&=& \{1+\dfrac{\Lambda}{3}f(t,n, \Lambda) \} S^{1}(0)\nonumber\\
S^{2}(t, n, \Lambda)&=& \{1- \dfrac{\Lambda}{3} f(t,n, \Lambda)\} S^{2}(0)
\end{eqnarray}
Also, for $ " \times " $ modes, we can find: 
\begin{eqnarray}\label{13}
S^{1}(t, n, \Lambda)&=& S^{1}(0)+ \dfrac{\Lambda}{3}f(t,n, \Lambda)S^{2}(0)\nonumber\\
S^{2}(t, n, \Lambda)&=&S^{2}(0)+  \dfrac{\Lambda}{3} f(t,n, \Lambda) S^{1}(0)
\end{eqnarray}
where $ f(t,n, \Lambda) $ by using $ D_{+,\times}(0,t) $  will be: 
\begin{eqnarray}
f(t,n, \Lambda)&=&1+\dfrac{3}{2\Lambda}  e^{i n t}  \frac{\left(3 n^2+1\right) \cos t+i n \left(n^2+3\right) \sin t}{ \left(n^2-1\right)} \nonumber\\&&~~~~~~~~~~~~~~~+  \sqrt{\frac{3}{\Lambda }} e^{i n t} \tanh({\sqrt{\dfrac{\Lambda}{3}}}t) \frac{\left(n^2+1\right) \sin t-2 i n \cos t }{n^2-1}
\end{eqnarray}
Note again that $ n$ is the wave number and should be a discrete number. In this case, the effect of gravitational waves on massless particles depends on $ n $ and $ \Lambda $. So, the movement of particles will be different from the flat case.\\
By comparing (\ref{12}) and (\ref{13}) with Relations (7.114) and (7.115) in Ref.\cite{6}, we can define $ h(t,n,\Lambda) $ and $ \omega(t,n,\Lambda) $ based on the real and imaginary parts of $f(t,n,\Lambda) $ as:
\begin{eqnarray}
\dfrac{\Lambda}{3} f(t,n,\Lambda)= h(t,n,\Lambda) e^{i \omega (t,n,\Lambda) t}
\end{eqnarray}
\begin{figure} 
	\subfloat[{}]{
		\includegraphics[width=0.45\textwidth]{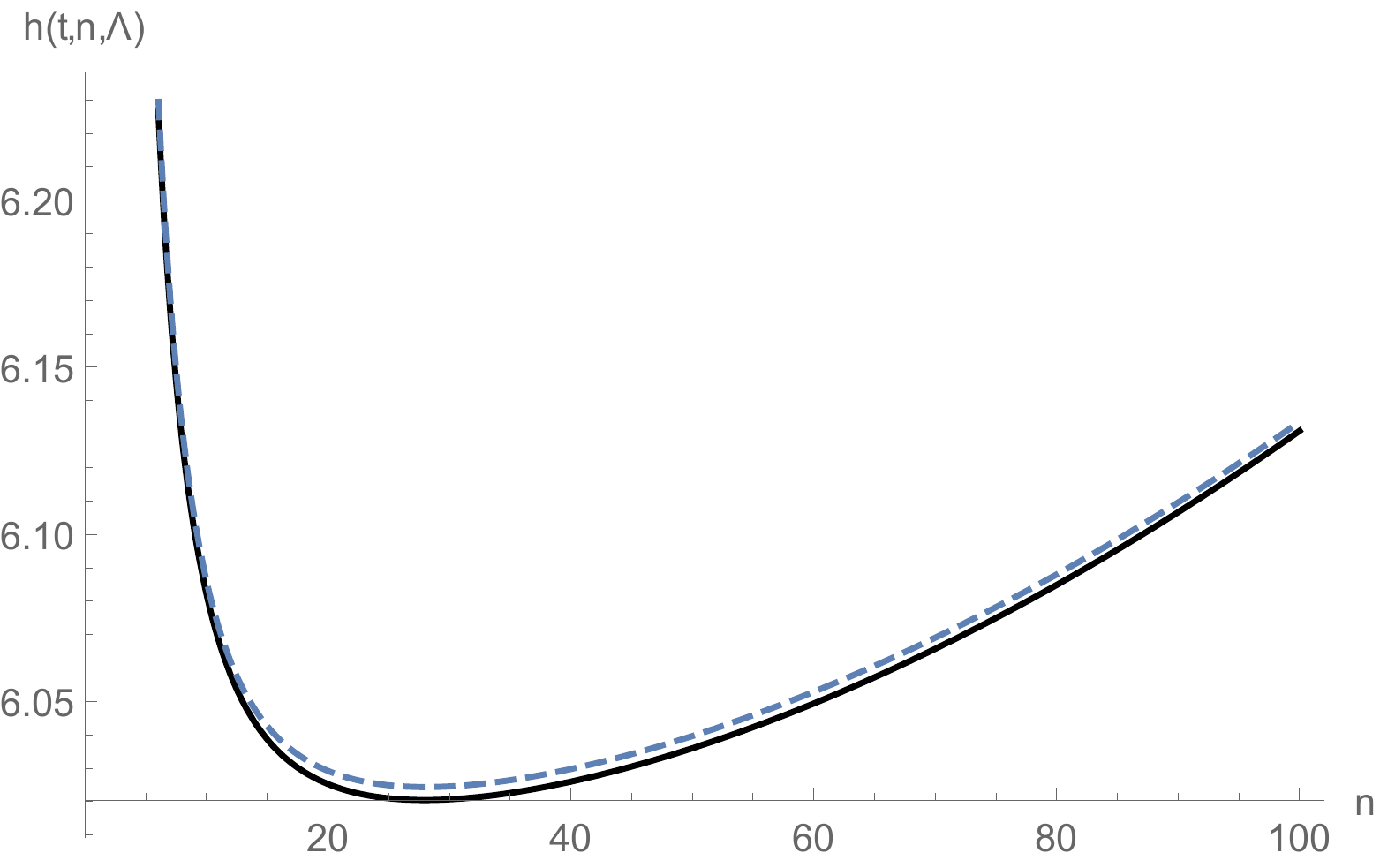} 
	} 
	\hfill 
	\subfloat[{}]{ 
		\includegraphics[width=0.45\textwidth]{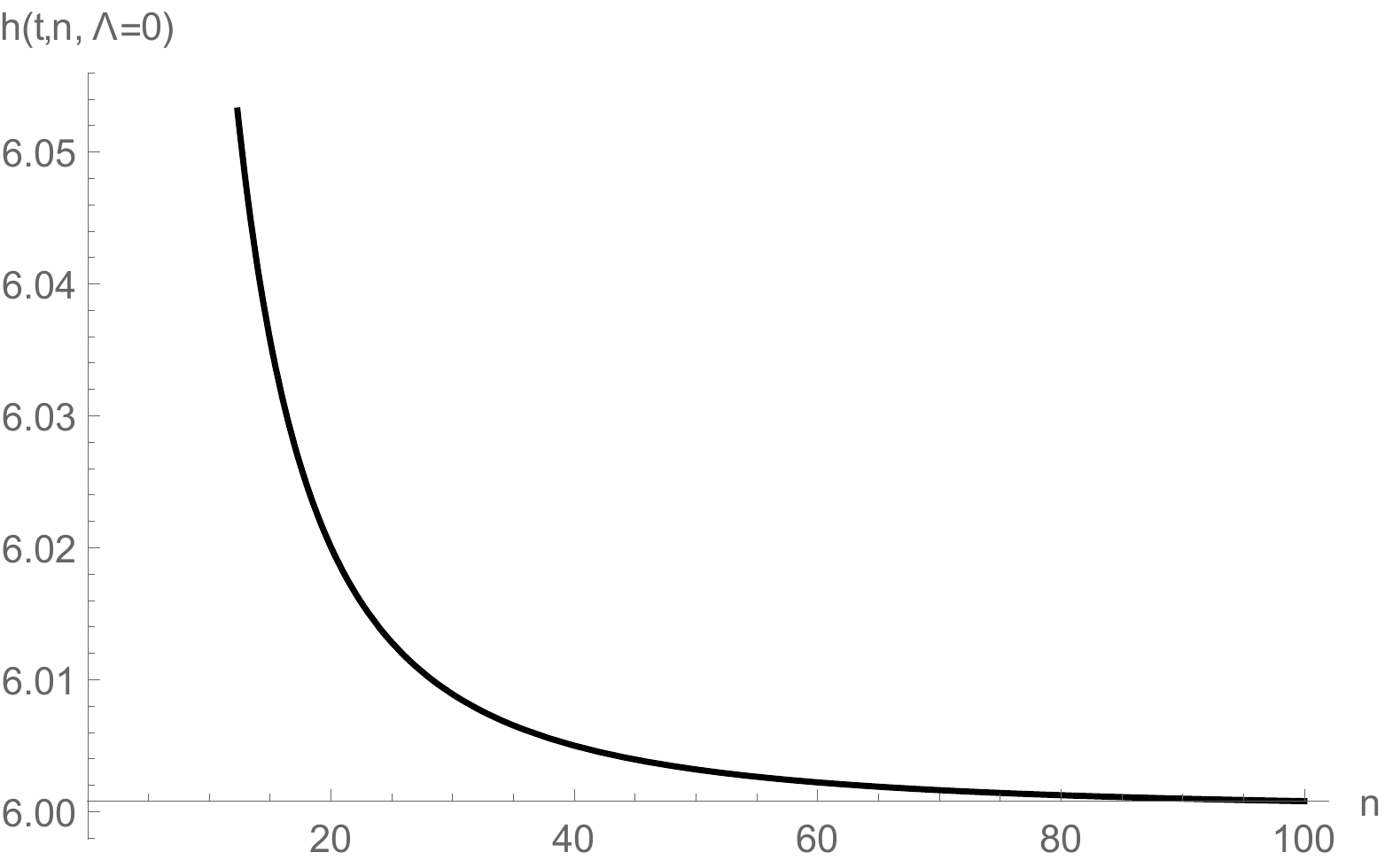} 
	} 
	\caption{  The plot of $ h(t,n,\Lambda) $ respect to $ n $  on two different ranges of time $ t $. (a): In the short time (or $ \lambda\gg L_{B}$), the amplitude has a minimum and increases with the large number of $ n $ (solid line). For $ \Lambda $ greater than 0.001, one can see its effect and the increase in the amplitude (dashed line is for $ \Lambda=0.003 $). ~~~~(b): In the long-time propagation, or $ \lambda\ll L_{B}$,  the amplitude reduces to the constant value (solid line). Comparing (a) and (b) shows that, for a small value of n, amplitudes behave like each other and are independent of the time of propagation.}  
	\label{h-n}
\end{figure}
where $ h(t,n,\Lambda) $ and $ \omega(t,n,\Lambda) $ are the amplitude and frequency (or phase term) of gravitational waves in closed spacetime, respectively, as:
\begin{eqnarray}
h(t,n,\Lambda)&=&\dfrac{2}{3(n^{2}-1)}\{[2 \Lambda \left(n^2-1\right)-3 n (n^2+3) \sin (t) \sin (n t)+3 (3 n^2+1) \cos (t) \cos (n t) \nonumber\\ &+& 2\sqrt{3\Lambda}\tanh (\sqrt{\dfrac{\Lambda}{3}}t) \left((n^2+1) \sin (t) \cos (n t)+2 n \cos (t) \sin (n t)\right)]^{2}\nonumber\\&+&[3 n \left(n^2+3\right) \sin (t) \cos (n t) + 3 \left(3 n^2+1\right) \cos (t) \sin (n t)\nonumber\\&+& 2\sqrt{3\Lambda} \tanh (\sqrt{\dfrac{\Lambda}{3}}t) \left((n^2+1) \sin (t) \sin (n t)-2 n \cos (t) \cos (n t)\right)]^{2}\}^{1/2}
\end{eqnarray}
and
\begin{eqnarray}
\omega(t,n,\Lambda)&=&\dfrac{1}{t} Arctan \{[3 n \left(n^2+3\right) \sin (t) \cos (n t)+3 \left(3 n^2+1\right) \cos (t) \sin (n t)\nonumber\\&+& 2\sqrt{3\Lambda}\left(3 n(n^2+3) \sin (t) \cos (n t)+3(3 n^2+1) \cos (t) \sin (n t) \right) \tanh (\sqrt{\dfrac{\Lambda}{3}}t)] \nonumber\\
&\times&[2 \Lambda (n^2-1)-3 n (n^2+3) \sin (t) \sin (n t)+3 (3 n^2+1) \cos (t) \cos (n t)  \nonumber\\ &+&2\sqrt{3\Lambda} \left( \left(n^2+1\right) \sin (t) \cos (n t)+2 n \cos (t) \sin (n t)\right) \tanh (\sqrt{\dfrac{\Lambda}{3}}t)]^{-1}\}
\end{eqnarray}
The amplitude of the gravitational wave depends on the time duration of the wave propagation. As usual, there are two widely separated spatio-temporal scales for the GWs\cite{Isac1, Isac2}, see also \cite{Marochnik:2012qe,Marochnik:2017tfa} for GWs effect on the second order perturbation of background metric and nonlinear effect such as GW turbulence \cite{Galt}. Let us define $L_{B}$  and $\lambda$  as  the length scale of variation of the background and  the scale of the ripple, respectively. In a short time $ \lambda\gg L_{B}$ and in a long time $ \lambda\ll L_{B}$ which is so-called " short wave approximation".
\\ In the short time (or $ \lambda\gg L_{B}$), amplitude has a minimum value and will increase for the large values of $ n $ and the effect of $ \Lambda $ is turned on from $ \Lambda \gtrsim 0.001 $. However, for the long-time propagation or short-wave approximation, it goes to a finite amount for a large value of $ n $, which is similar to the flat case. The effect of $ \Lambda $’s can be inferred from $ \Lambda\gtrsim 0.0001 $ (Fig.\ref{h-n}). 
\begin{figure}
	\includegraphics[scale=0.7]{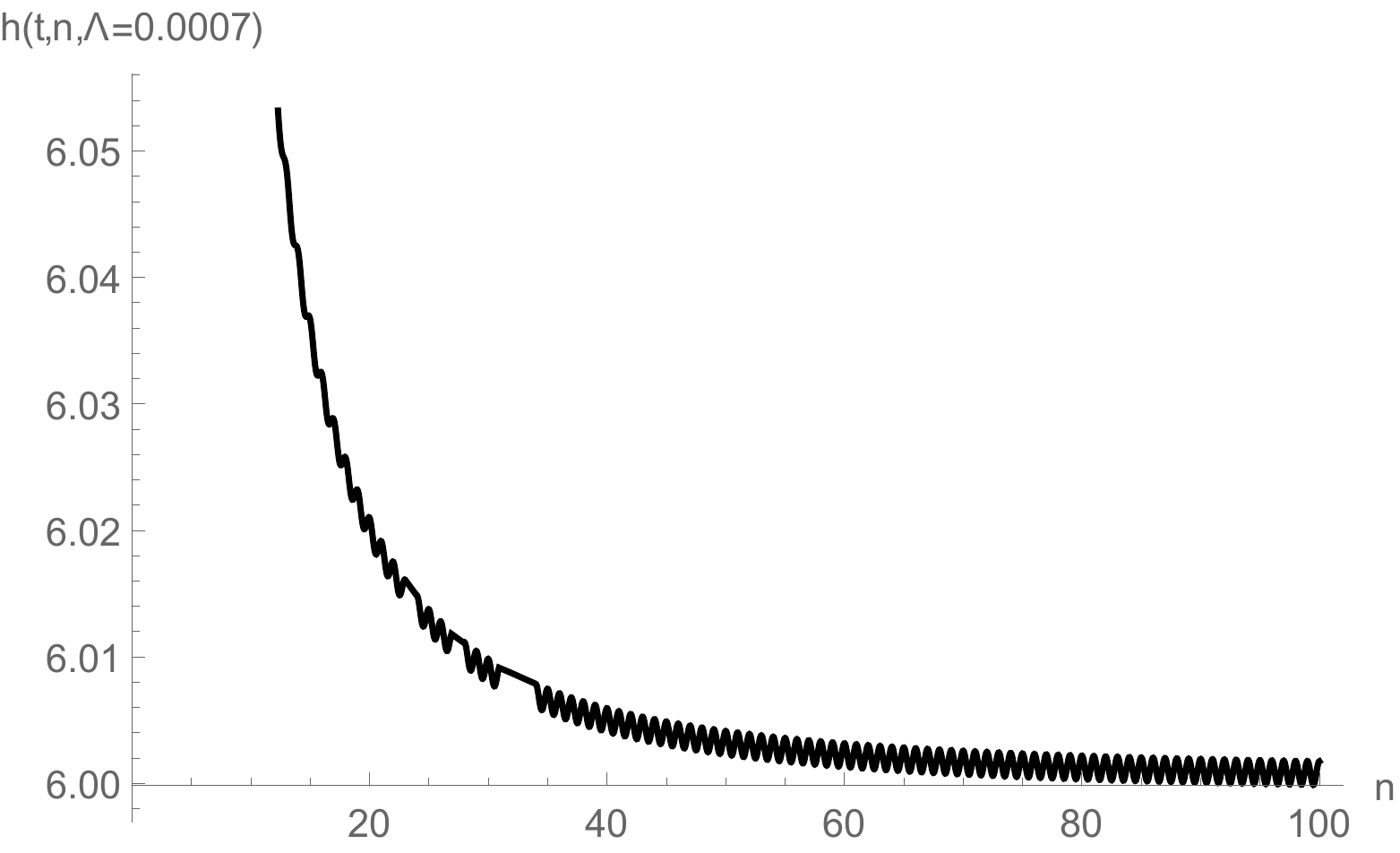}
	\caption{ The rate of $ h (t,n,\Lambda=0.0007)$ with respect to $ n $ for the long-time propagation or short wave approximation. It is noteworthy that $ \Lambda $ induces the oscillating behavior to the amplitude. This effect is visible at the amplitude of greater than 0.0001.}
	\label{h-lambda}
\end{figure}
For the small value of $ n $, its dynamic relation to $ n $ is independent of the time of wave propagation.\\ 
The role of $ \Lambda $ is different in each case: for the short time, $ \Lambda $’s effect increases the amplitude (Fig.\ref{h-n}, (a) dashed line), while for the long time, the effect of $ \Lambda $ makes an oscillatory behavior around the state of $ \Lambda=0 $ (Fig.\ref{h-lambda}).\\ In the Planck result \cite{Planck2}, the $ \Lambda $ is $ (4.24\pm 0.11) \times 10^{-66} eV^{2}= (2.846\pm 0.076 ) \times 10^{-122} m_{Pl}^{2}\approx 10^{-56} cm^{-2}$ in natural units, where $ m_{Pl} $ is the Planck mass. Therefore, to make cosmological constant effective for GW’s, it should be approximately 52 orders of magnitude larger than the present observable value. Here, $ \rho_{\Lambda} $ should be of order $ \rho_{_{Electro-Weak}}\sim 10^{23} g~cm^{-3} $, where it is the stage of softly broken super-symmetric in the universe \cite{Polch}.
\\ For a small $ n $, the frequency can be written in a series of time as:
\begin{eqnarray}
w(t,n,\Lambda)= n + \dfrac{3 \tan(t) }{t}+\dfrac{2 \Lambda \left(-5 t-t \sin ^2(t)+6 \sin (t)+t \cos ^2(t)+2 t \cos (t)\right)}{3 t (\cos (2 t)+1)}\,.
\end{eqnarray}
In the long-time propagation, the second and third terms can be ignored. But, for a short time, its rate is similar to the previous one ( Fig.\ref{w-n}), so the frequency is independent of the time propagation of the GWs. From $ \Lambda\geq 0.1 $, it will affect the decrease in the frequency of gravitational waves.
\begin{figure}
	\includegraphics[scale=0.7]{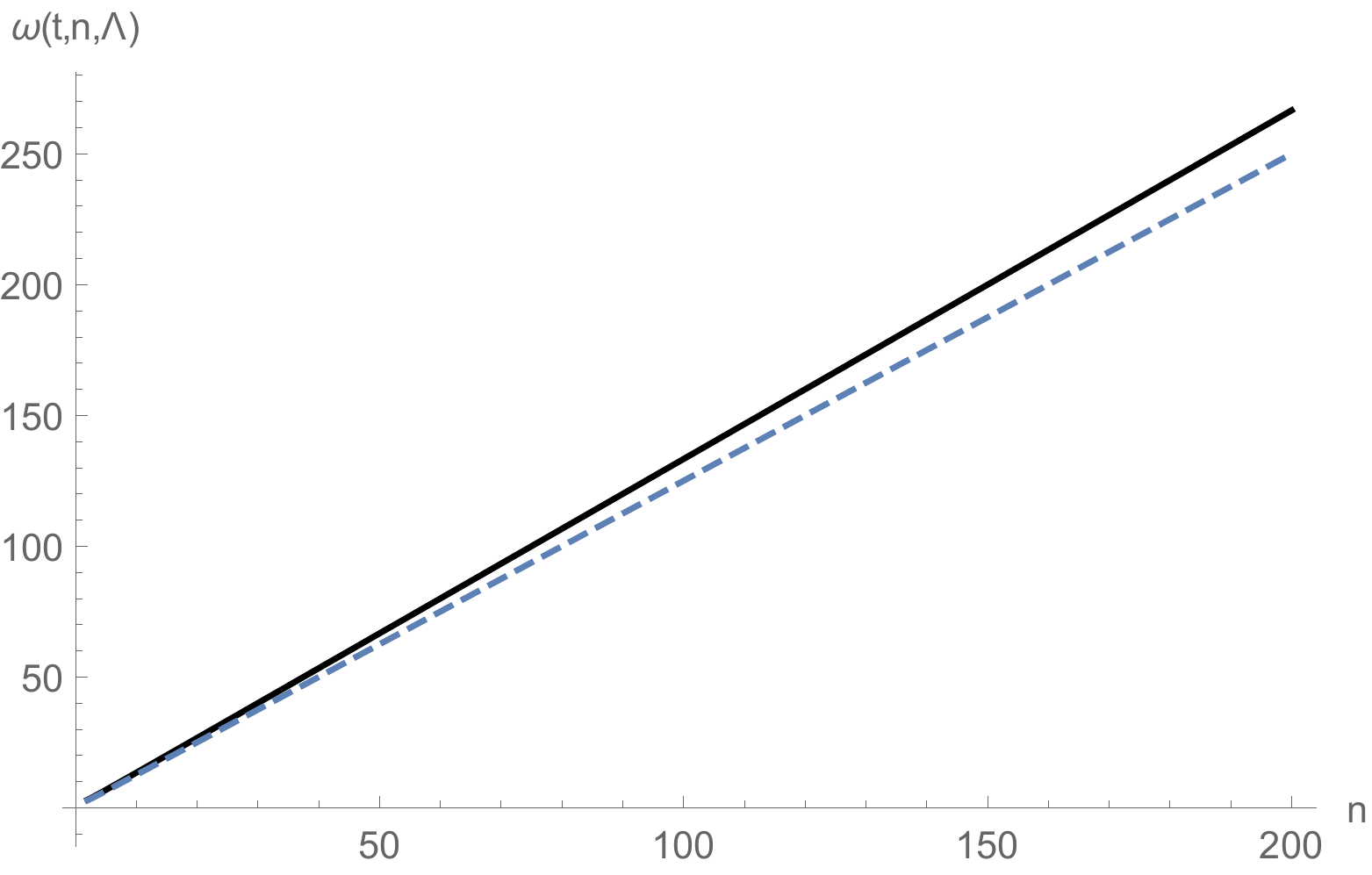}
	\caption{The rate of $ w (t,n,\Lambda)$ with respect to $ n $ for the short time propagation; the effect of $ \Lambda $ appears when exceeding $ 0.1 $ (see dashed line).}
	\label{w-n}
\end{figure} 
In the short-time propagation, the amplitude and frequency behaviors are very similar to each other although the amount of $ \Lambda $’s effect on them is different (Figs.\ref{h-n}, .\ref{h-lambda} and .\ref{w-n}).
\\
Now, we are looking for the effect of exact solutions on the nearby particles. So, Relations (\ref{12}) and (\ref{13}) become: 
\begin{eqnarray}
S^{1}(t, n, \Lambda)&=& \{1+\dfrac{\Lambda}{3}+\dfrac{1}{2}  e^{i n t}  \frac{\left(3 n^2+1\right) \cos t+i n \left(n^2+3\right) \sin t}{ \left(n^2-1\right)} \nonumber\\&&~~~~~~~~~~~~~~~+  \sqrt{\frac{\Lambda}{3 }} e^{i n t} \tanh
(\sqrt{\dfrac{\Lambda}{3}}t) \frac{\left(n^2+1\right) \sin t-2 i n \cos t }{n^2-1} \} S^{1}(0)\nonumber\\
S^{2}(t, n, \Lambda)&=& \{1-\dfrac{\Lambda}{3}-\dfrac{1}{2}  e^{i n t}  \frac{\left(3 n^2+1\right) \cos t+i n \left(n^2+3\right) \sin t}{ \left(n^2-1\right)} \nonumber\\&&~~~~~~~~~~~~~~~-  \sqrt{\frac{\Lambda}{3 }} e^{i n t} \tanh
(\sqrt{\dfrac{\Lambda}{3}}t) \frac{\left(n^2+1\right) \sin t-2 i n \cos t}{n^2-1}\} S^{2}(0)
\end{eqnarray}
\begin{figure}
	\includegraphics[scale=0.28]{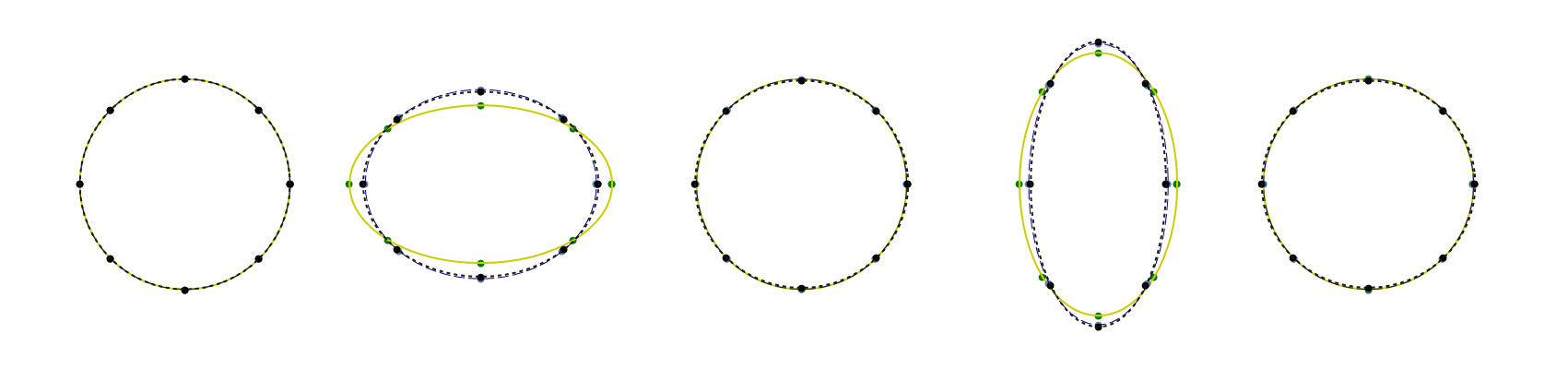}
	\caption{The effect of the gravitational wave with the wave number n=2 in closed with $ \Lambda=0 $ (dotted blue line), $ \Lambda=0.1 $ (dotted-black), and flat spacetime (solid-yellow line) on the nearby particles; the effect of the cosmological constant is the oscillatory effect.}
	\label{fig:n=2}
\end{figure}
By using numerical methods we compare our result with the flat case  which it only examines the frequency and mode of the polarization. Obviously, $ n=1 $ is not the answer. 
For $ n=2 $, moving the nearby particles is shown in Fig.\ref{fig:n=2}. At lower $n $'s, the difference between the flat and closed cases is quite evident with $ \Lambda=0 $ (dotted-blue). If we assume $ \Lambda=0.1 $ or larger, it has a very small change on the frequency of particles ring (Fig. \ref{fig:n=2}, dotted-black). It can be predicted that, if the cosmological constant is introduced as a source term, only the lower frequency modes of GWs will be affected by the background and its detection can be used to analyze the Dark energy \cite{Ivan1, Nowa}. In comparison, if $ \Lambda $ has the role of isotropic motion in the geodesic deviation equation, the ring of particles is deformed into an ellipse and their normal modes have $\dfrac{\pi}{4}$ relative to each other\cite{Bicak3}.
For large $ n $, expanding the universe destroys the oscillation of gravitational waves (Fig.\ref{fig:n=20}). These results were obtained for $ \times $ mode. Thus, the effect of GWs in the closed background shows itself only in low wave numbers. In the following, we will show the effect of closed GWs on the nearby particles and the quantitative difference between the closed and flat GWs. 
\section{  Effect of Closed Gravitational Wave on Nearby Particles   }
The Hubble parameter, $H(t)=\dfrac{\dot{a}(t)}{a(t)} $, is the  effect of the expanding universe; its impact on massive objects has been seen on the scale of galaxies and clusters \cite{Rosh}.
Yet, the source of GWs is located at the cosmologically large distances from the observer, e.g. $ z\geq 0.1 $, the effect of cosmological expansion on GWs are relevant and the GW150914 ($ z=0.09 $) is in this limit. In the wide range of the studies taken into account, the expansion of background \cite{49,50,51,52,53,54,55}. Since the test particles are very close to each other and move very slowly, only their relative motion has been taken into consideration. Therefore, neither the first term, nor the Hubble constant, in the Eq.\eqref{deviation}, has any effect on this circle of nearby particles. On the other, the first and second terms of Eq. (\ref{deviation}) are very smaller than the last term and we can ignore them. For slowly-moving particles, we have $ \tau=x_{0}=t $ and the geodesic deviation equation becomes:
\begin{figure}
	\includegraphics[scale=0.28]{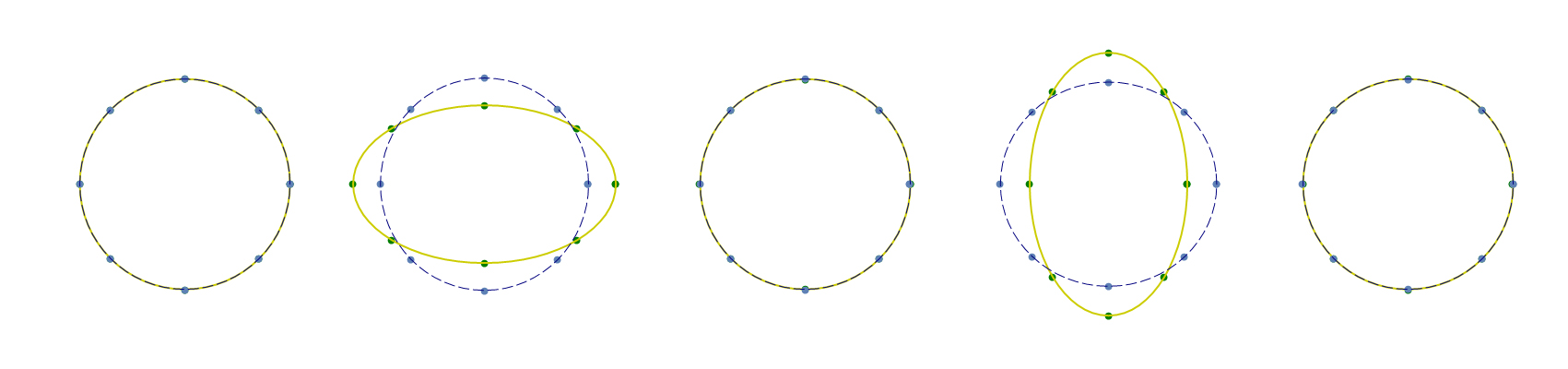}
	\caption{The effect of the gravitational wave with the wave number n=20  in closed with  $ \Lambda=0.1 $ (dotted-black) and flat spacetime (solid-yellow line) on the nearby particles. The GWs will not fluctuate the massless particles.}
	\label{fig:n=20}
\end{figure}
\begin{eqnarray}\label{27}
\dfrac{\partial^{2}}{\partial t^{2}}S^{i}=\dfrac{1}{2} S^{j}\dfrac{\partial^{2}}{\partial t^{2}} D^{i}_{~j}
\end{eqnarray}
As mentioned, $D^{i}_{~j}$ is the solution for gravitational waves equation in curved space time \cite{Gholizadeh}. 
By using (\ref{29}), the deviation equations for $ + $ mode will be:
\begin{eqnarray}\label{11}
\frac{d^{2}}{d t^{2}}S^{1} &=&\frac{1}{2}S^{1}\frac{d^{2}}{d t^{2}} D^{1}_{~1}+\frac{1}{2}S^{2}\frac{d^{2}}{d t^{2}} D^{1}_{~2}+\frac{1}{2}S^{3}\frac{d^{2}}{d t^{2}} D^{1}_{~3} \nonumber\\
\frac{d^{2}}{d t^{2}}S^{2} &=&\frac{1}{2}S^{1}\frac{d^{2}}{d t^{2}} D^{2}_{~1}+\frac{1}{2}S^{2}\frac{d^{2}}{d t^{2}} D^{2}_{~2}+\frac{1}{2}S^{3}\frac{d^{2}}{d t^{2}} D^{2}_{~3}\nonumber\\
\frac{d^{2}}{d t^{2}}S^{3}&=&\frac{1}{2}S^{1}\frac{d^{2}}{d t^{2}} D^{3}_{~1}+\frac{1}{2}S^{2}\frac{d^{2}}{d t^{2}} D^{3}_{~2}+\frac{1}{2}S^{3}\frac{d^{2}}{d t^{2}} D^{3}_{~3}
\end{eqnarray}
Assuming that the particles remain on the $ S_{1}-S_{2} $ plane, we will have $ S_{3}=0 $ and
\begin{eqnarray}
D_{+,\times}(z=0,t) =\dfrac{\cos(t) \cos(nt) \pm n\sin(t) \sin(nt)}{1- n^{2}}
\end{eqnarray} 
We see that the gravitational wave functions are also dependent on wave numbers. For $n=0$, the solution of the gravitational waves equation is:
\begin{equation}
S_{1,+}(S_{2,+})=c_{1} MathieuC(0,1,\frac{t}{2})\pm c_{2}MathieuS(0,1,\frac{t}{2})
\end{equation}
where in general Mathieu C(a, q, z) and Mathieu S(a, q, z) are even and odd solutions of the Mathieu differential equation, respectively  \cite{abramowitz}. As seen in Fig.\ref{fig:3}, this solution is not stable. Also, $n=1$ is a gauge mode, the solution of which is $x(t)=constant $. As seen in Fig. \ref{fig:3}, the solution is not an oscillation, so that it has no effect on the particles. For $ n=2 $, there is no analytical solution; but, for mode $ n=3 $, the solution of Eq.(\ref{27}) can be written as the Heun C function\cite{Hortacsu} as:
\begin{figure}
	\centering
	\includegraphics[scale=0.7]{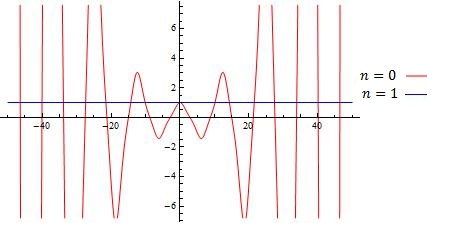}
	\caption{The solution for n=0 (redline) as a Mathieu function which is not stable. The solution n=1 (blue line) is a gauge mode which has no oscillation. }
	\label{fig:3}
\end{figure}
\begin{eqnarray}
S_{1,+}(t)=(c_{1}+ c_{2}\cos(t))e^{\sqrt{2}\cos^{2}(t)}HeunC(2\sqrt{2},\frac{1}{2},\frac{1}{2},\frac{1}{4},0,\cos^{2}(t))
\nonumber\\
S_{2,+}(t)=(c_{3}- c_{4}\cos(t))e^{\sqrt{2}\cos^{2}(t)}HeunC(2\sqrt{2},\frac{1}{2},\frac{1}{2},\frac{1}{4},0,\cos^{2}(t))
\end{eqnarray}
where the HeunC function is the solution of the Heun Confluent equation and  $c_{1}$, $c_{2}$, $c_{3}$ and $c_{4}$ are constant. For other modes, there are no analytical solution. \\
Even though these equations have no exact analytical solutions for any values of $n $, the most simple solutions are similar to (7.114) and (7.115) of ref \cite{6}, except that the time evolution of gravitational waves in closed spacetime is different from that of the flat case (i.e. $ D_{+,\times}(z=0,t) $ has been used instead of $ \exp(iwt) $ in ordinary GW's equation). Therefore, the terms of $ O(x^{3})$ and high orders are ignored and the solutions can be written in terms of the series expansion for any value of the wave number as:
\begin{eqnarray}\label{plus}
S_{1,+} (n,t)&=& \{1+\sum_{i=2}^{\infty} g_{i}(n)t^{i}\}S_{1}(0) \nonumber\\
S_{2,+} (n,t)&=&\{1-\sum_{i=2}^{\infty} g_{i}(n)t^{i}\}S_{2}(0)
\end{eqnarray}
where  $ \sum_{i=2}^{\infty} g_{i}(n)t^{i}= f(t, n, \Lambda=0) $ and  $\times$ mode gives:
\begin{eqnarray}
S_{1,\times}(n,t) &=& S_{1}(0)+ \sum_{i=2}^{\infty} g_{i}(n)t^{i}S_{2}(0) \nonumber\\
S_{2,\times}(n,t) &=& S_{2}(0)+ \sum_{i=2}^{\infty}  g_{i}(n)t^{i}S_{1}(0)
\end{eqnarray}
\begin{figure}
	\centering
	\includegraphics[scale=0.5]{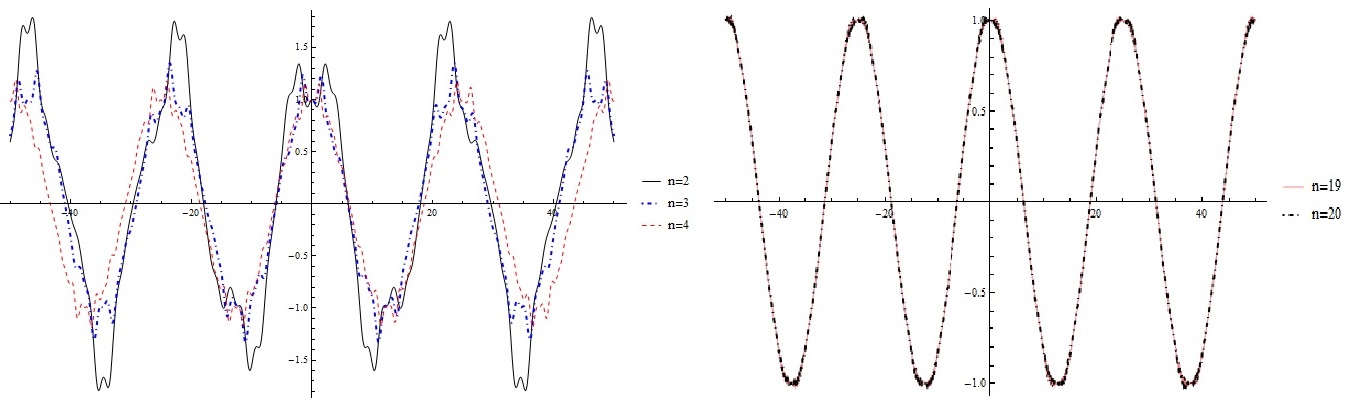}
	\caption{Left: The solution of equation for n=2(black), n=3(dot-dashed) and n=4(dashed). As seen, there is difference between the lower mode solution of the closed background and the flat case. Right: The solution of equation for n=19(solid line) and n=20(dot-dashed) and higher. In the upper modes, the solutions will be the same as the flat case.}
	\label{fig:3000}
\end{figure}
The particles are located on the circle of radius $ S_{1}(0)=S_{2}(0)=1 $ before getting hit by GWs.  As mentioned, in the $ n=0 $, the solution is not stable and $ n=1 $ is a gauge mode which is not an oscillation. For the lower modes (Fig. \ref{fig:3000} Left), these solutions differ from the flat cases. Then, the amplitude of particles oscillation is different from that of the flat case, as we can see in Fig.\ref{fig1} and Fig.\ref{fig2}. The interesting point is that there is a new oscillation in every period; in other words, the particles oscillation in each period is different from that of others. When the mode number increases (for example, $n=18$ ,$n=19$ and $n=20$ or higher wave numbers), the solution in the closed background will be the same as the flat spacetime; these two cases cannot be distinguished ( Fig. \ref{fig:3000} Right and Fig.\ref{fig4} ). When the particles oscillate like the flat spacetime, this ensures the stability of the solutions. 
\subsection{Closed Background Effect}
Now, we consider the next order terms of $O(x^{3})$, that is all the polarization matrix sentences of the (\ref{29}). So, the solution of Eq.(\ref{11}) with the condition of the remaining particles in $ S_{1}-S_{2} $ and $ S_{3}=0 $ for $ + $ mode becomes:
\begin{eqnarray}
S_{1,+} &=& \{1+\sum_{i=2}^{\infty} k_{i}(n)t^{i}\}S_{1}(0) \nonumber\\
S_{2,+} &=&\{1-\sum_{i=2}^{\infty} k_{i}(n)t^{i}\}S_{2}(0)
\end{eqnarray}
And the $\times$ mode solutions is:
\begin{figure}
	\includegraphics[scale=0.47]{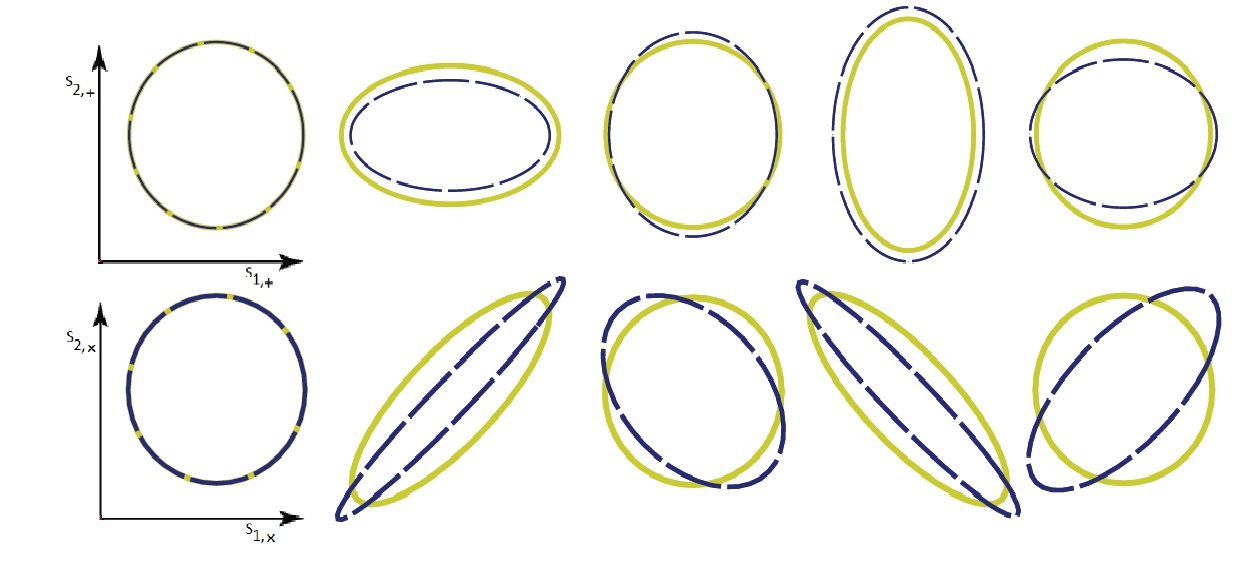}
	\caption{The effect of the gravitational wave with the wave number n=2  in closed (dashed line ) and flat spacetime (solid line) on the nearby particles. In the next periods in closed spacetime, unlike the flat case, these shapes of oscillation cannot be seen and there are new shapes.}
	\label{fig1}
\end{figure}
\begin{eqnarray}
S_{1,\times} &=& S_{1}(0)+ \sum_{i=2}^{\infty} h_{i}(n)t^{i}S_{2}(0) \nonumber\\
S_{2,\times} &=& S_{2}(0)+ \sum_{i=2}^{\infty} v_{i}(n)t^{i}S_{1}(0)
\end{eqnarray}
It can be seen that the solutions are different for each mode, e.g.  $ k_{i}(n) $ is different from $ h_{i}(n) $  and $v_{i}(n)$. There is also asymmetry in the solution of mode $ \times $, because of the difference of $ h_{i}(n) $ from $ v_{i}(n)$. 
With these solutions, the particles start to oscillate at first. But, their oscillation will not remain stable later, i.e. the amplitude and frequency of particles will change with time in an inconsistent way for each wave number. By defining $ S_{\times}=\sqrt{S_{1,\times}^{2}+S_{2,\times}^{2}} $, the movements of test particles will not be stable, as shown in Fig. \ref{fig3}, for lower wave numbers. In high wave numbers, the particles will not be oscillating. This is true for $+$ mode as well, i.e. the closed structure operates as an anomaly effect for harmonic oscillation.
\begin{figure}
	\includegraphics[scale=0.47]{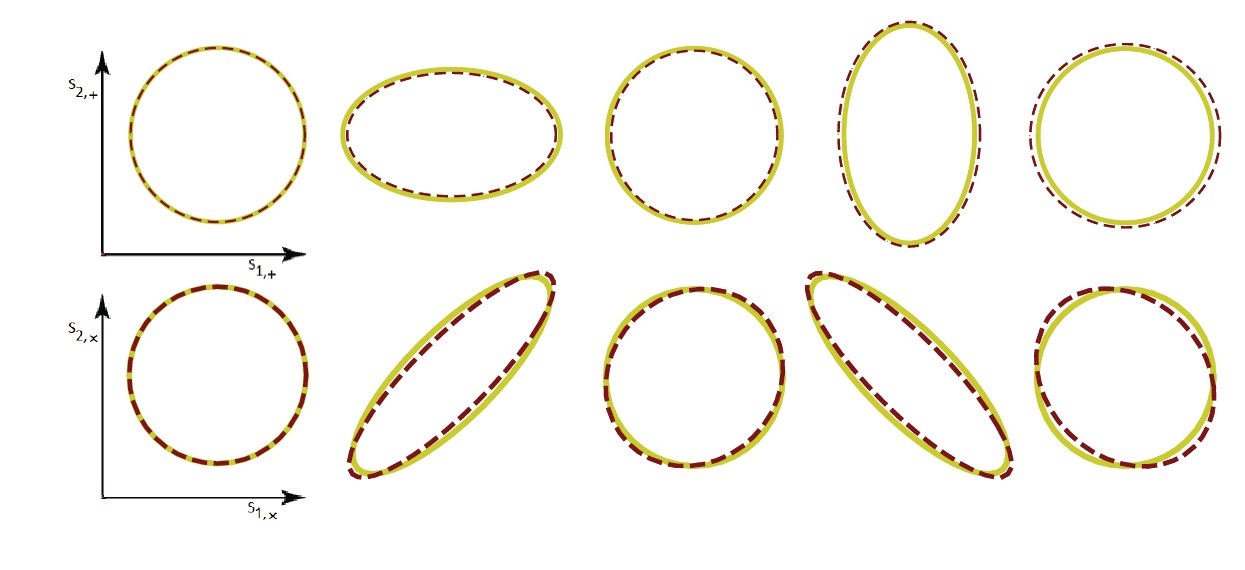}
	\caption{The effect of the gravitational wave with the wave number n=3  in closed (dashed line ) and flat spacetime (solid line) on the nearby particles. In the next periods in closed spacetime, the above shape will not be repeated.}
	\label{fig2}
\end{figure}
 \section{Quantitative Difference between Closed and Flat GWs }
For quantitative comparison of our results with the flat case, we should define:    
 \begin{eqnarray}
 \Delta S^{i}(t)=S^{i}_{closed}(t)-S^{i}_{flat}(t)
 \end{eqnarray}
where $ S^{i}_{flat}(t)= \dfrac{1}{2} h e^{i^{w_{_{GW}}t}}$ and $ h $ and $ w_{GW} $ are amplitude and frequency of GWs in flat spacetime, respectively. So,
 \begin{eqnarray}
 \Delta S^{i}(t)=[\dfrac{\Lambda}{3}f(t,n, \Lambda)-\dfrac{1}{2}h e^{iw_{GW}t}]\delta^{ij} S^{j}(0) ~~~~~,~~~~~i=1,2
 \end{eqnarray}
for $ "+ "$ mode $ i=j $ . This term, $ \dfrac{\Lambda}{3}f(t,n, \Lambda)-\dfrac{1}{2}h e^{iw_{GW}t} $ could separate the effect of two cases of flat and closed approaches. Therefore,     
 \begin{eqnarray}
 \dfrac{\Delta S^{i}(t)}{S^{i}(0)}&=&\dfrac{\Lambda}{3}f(t,n, \Lambda)-\dfrac{1}{2}h e^{iw_{GW}t}\nonumber\\
 &\cong&\dfrac{\Lambda}{3}+\dfrac{3n^{2}+1}{2(n^{2}-1)} \cos(t) \cos(nt) - \dfrac{n(n^{2}+1)}{2(n^{2}-1)} \sin(t) \sin(nt)-\dfrac{1}{2} h(f) \cos(w_{gw} t)\nonumber\\
 \end{eqnarray}
 \begin{figure}
 	\includegraphics[scale=0.45]{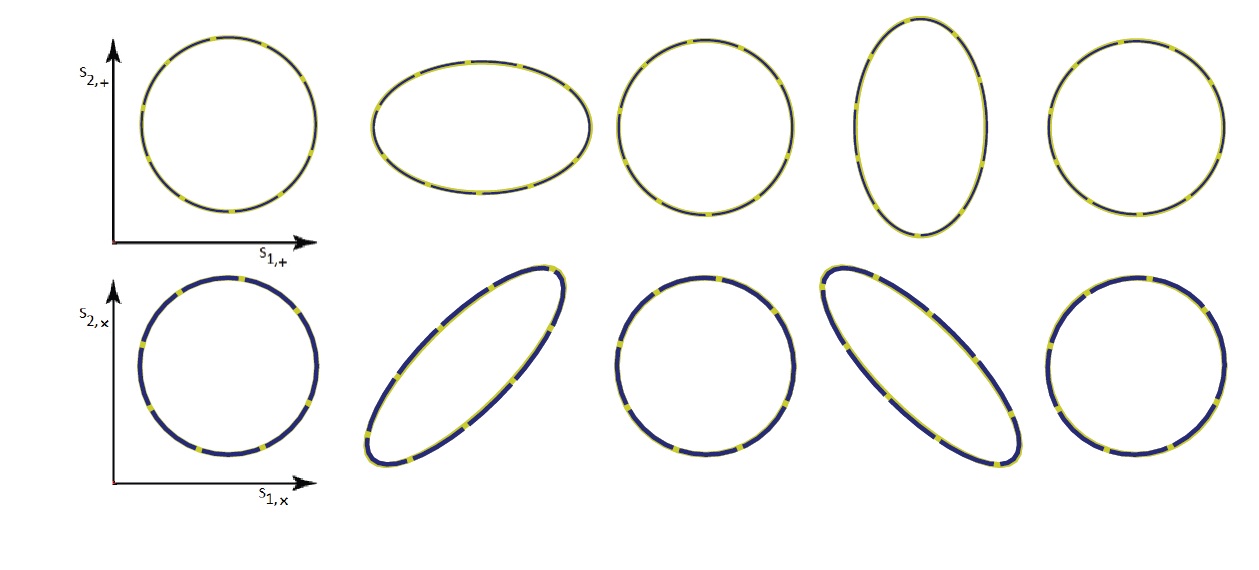}
 	\caption{ The effect of the gravitational wave with the wave number n=20  in closed (dashed line ) and flat spacetime (solid line) on the nearby particles. In high wave numbers, the effects of closed and flat background are the same.}
 	\label{fig4}
 \end{figure}
 for $ "+ "$ mode $ i=j $ . This term, $ \dfrac{\Lambda}{3}f(t,n, \Lambda)-\dfrac{1}{2}h e^{iw_{GW}t} $ could separate the effect of two cases of flat and closed approaches. Therefore, 
 This relation can be interpreted as the sensitivity of detector for a land-based detector, such as LIGO or VIRGO. We can explore the differences between the two cases in the same interval time and duration of the gravitational wave passes. Here, $ t $ can be considered as the length of time, in which the detector is turned on. Therefore, for instance, considering the $ \Lambda $ at the moment \cite{Planck2}, amplitude spectral density, $ 10^{-23} < h(f)< 10^{-19} $, and observed frequency, $ 10 (Hz) < f < 10^{4} (Hz) $ from the first detection of GWs, GW150914 \cite{Ligo}, one can see that the amount of sensitivity depends on the $ \Lambda $ and $ n $.   
 \begin{figure}
 	\includegraphics[scale=0.5]{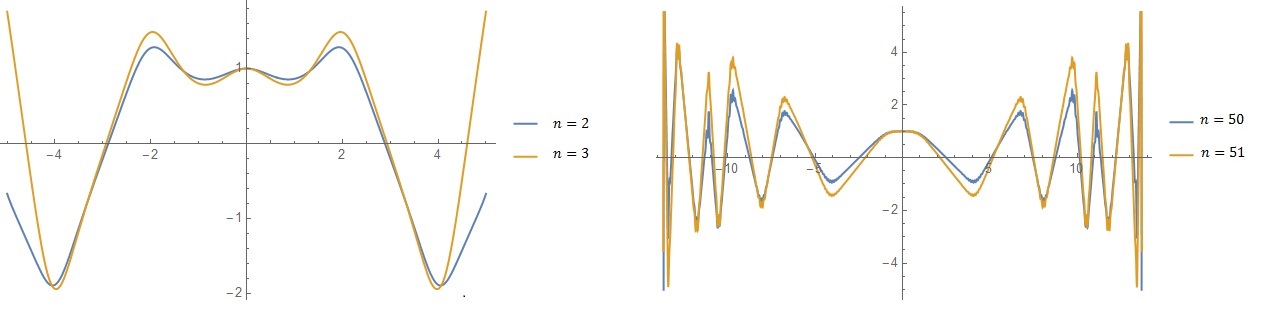}
 	\caption{ The plot of $ S_{\times}=\sqrt{S_{1,\times}^{2}+S_{2,\times}^{2}} $ for the nearby particles concerning the time in closed spacetime in the presence of the $ O(x^3)$ terms for lower and higher wave numbers. As seen in the next times, in lower wave numbers, both the frequency and amplitude will change momentarily and the system of particles will not remain stable. Also, in high wave numbers, the particles will not be oscillating.}
 	\label{fig3}
 \end{figure}
\section{Discussion and Conclusion}
So far, no study has been conducted on the frequency and amplitude of GWs as well as their effect on the nearby particles in closed de Sitter spacetime.  The amplitude and frequency of gravitational waves are obtained by comparing the solutions of deviation equation in closed background, with its flat case independent of its $+ $ or $ \times$ modes. The behavior of the amplitude relative to the wave number depends on the time duration of wave propagation, while the frequency is independent. The effect of the cosmological constant on GWs in closed spacetime is seen nearly after the epoch of Electro-Weak in the universe. Its effect on amplitude is weaker than on the frequency, because if $ \Lambda \geq 0.0001 $, the amplitude is affected. The same happens for frequency when $ \Lambda \geq 0.1 $. Also, the role of $ \Lambda $ is different for them; it reduces the frequency, while its effect on the amplitude is oscillatory with the dissipation effect. \\
The effect of this wave, which is on the large scale, can be used for the nearby particles in the small scale, because it is the solution of vacuum field equation of tensor perturbation and its traceless-transverse properties. Nevertheless, the effect on the nearby particles can only be seen at lower wave numbers, while in large numbers, the expansion of the universe will eliminate the oscillatory effect of GWs.\\
The second approach corresponds to the flat case, but the time evolution of GWs in de Sitter spacetime differs from the Minkowski background. It has been concluded that the behavior of the low-frequency modes will be affected by the de-Sitter background. The differences between the two cases are also seen in the amplitude of the particle oscillations. Although the infrared modes are sensitive to the cosmological constant, it was found that the propagation of the gravitational waves would be observed at modes lower than the infrared ones when the wavelength of gravitational waves is smaller than the de Sitter radius of curvature and higher than the deviation. Therefore, the gravitational radiation can be used to test the intrinsic curvature of the large-scale spacetimes.\\ The particle oscillation of flat spacetime, which is recovered within the closed spacetime for the large wave numbers, indicates the presence of stabilities in the solutions. Additionally, in lower wave numbers within closed spacetime, the particles’ oscillation is not harmonic and, rather, have small fluctuations in the back-point or at the maximum amplitudes. In the next order terms ($ O(x^{3}) $ and higher), caused by the closed structure, the particles start to oscillate; but after a short time, the ring of the test particles will become unstable. Hence, with the new approach, we can reconfirm that de Sitter background space operates as the cut-off oscillation. \\
More interestingly, the only way to investigate the curvature of background spacetime by detecting gravitational waves is perhaps the direct study of GWs in lower wave numbers (very low frequencies) or at very high wavelengths. However, the precision of this detector must be in the order of $ \Lambda $-measurement at the moment in order to measure the distance between the two points, i.e. one under the influence of plane wave (flat wave) and the other under the influence of closed gravitational waves.
\appendix 
\section{Traceless-transverse Gauge or TT Gauge}
The transverse-traceless gauge in curved spacetime is defined as:  
	\begin{eqnarray}
	D_{i0}=0~~~~~~~,~~~~~~~D^{i}_{~i}=K x^{j}x^{k}D_{jk}~~~~~~~,~~~~~~~\partial_{i}D_{ik}=K x^{i}x^{j}\partial_{i}D_{jk}-4K x^{l}D_{kl} 
	\end{eqnarray}
	In the TT gauge, GWs have an specially simple form. If a test mass is at rest at $\tau=0$, from the geodesic equation, we have  $ \dfrac{dx^{i}}{d\tau}=0$, the mass initially has been taken at rest when $ \Gamma_{00}^{i}=0$. By writing   $ g_{\mu\nu}=\bar g_{\mu\nu}+ h_{\mu\nu}$ and expanding to first order in  $ h_{\mu\nu}$, the Christoffel symbol $ \Gamma_{00}^{i}$  ( from equations (3) and (11) in \cite{Gholizadeh}) becomes:
	\begin{eqnarray}
	\Gamma_{00}^{i}&=& \bar{\Gamma}_{00}^{i}+\delta  \Gamma_{00}^{i}\nonumber\\&=&\dfrac{\dot{a}}{a}\delta_{0i}+\dfrac{1}{2a^{2}}[2\dot{h}_{0i}-\partial_{i}h_{00}-2 K x^{i}x^{j}\partial_{0}h_{0j}+Kx^{i}x^{j}\partial_{j}h_{00}]
	\end{eqnarray}
	This quantity vanishes because both $h_{00}$ and $h_{0i}$ are set to zero. Therefore, if at time $\tau=0$, $ \dfrac{dx^{i}}{d\tau}$ and its derivative $ \dfrac{d^{2}x^{i}}{d\tau^{2}}$ vanish; so, $ \dfrac{dx^{i}}{d\tau}$ remains zero at all times. In the TT gauge, particles were at rest before the arrival of the wave and, then, remained at rest even after the arrival of the wave. Therefore, in the closed background, the GWs have all properties of transverse-traceless gauge.

\end{document}